\documentclass[final,1p,times,twocolumn]{elsarticle}
  
\usepackage{geometry}
\usepackage{graphicx}
\usepackage{amssymb}
\usepackage{verbatim}
\usepackage{lineno}
\usepackage{tabularx}
 
\usepackage{float} 
\usepackage {hyperref} 
\hypersetup{
colorlinks=true,
linkcolor=blue,
filecolor=magenta, 
urlcolor=cyan,
}

\date{}

\begin{document}

\begin{frontmatter}


\title{Secondary Use of Electronic Health Record: Opportunities and Challenges}


 
 \author[label1]{Shahid Munir Shah}
 \author[label1,label2]{Rizwan Ahmed Khan}
 
\address[label1]{Faculty of IT, Barrett Hodgson University, Karachi, Pakistan}
\address[label2]{LIRIS, Universit\`e de Lyon, France}


\begin{abstract}

In present technological era, healthcare providers generate huge amount of clinical data on daily basis. Generated clinical data is stored digitally in the form of Electronic Health Records (EHR) as a central data repository of hospitals.  Data contained in EHR is not only used for the patients' primary care but also for various secondary purposes such as clinical research, automated disease surveillance and clinical audits for quality enhancement.  Using EHR data for secondary purposes without consent or in some cases even with consent creates privacy issues for individuals. Secondly, EHR data is also made accessible to various stake holders including different government agencies at various geographical sites through wired or wireless networks. Sharing of EHR across multiples agencies makes it vulnerable to cyber attacks and also makes it difficult to implement strict privacy laws as in some cases data is shared with organization that is governed by specific regional law. Privacy of an individual could be severely affected when their sensitive private information contained in EHR is leaked or exposed to public. Data leak can cause financial losses or an individuals may encounter social boycott if their medical condition is exposed in public. To protect patients personal data from such threats, there exists different privacy regulations such as General Data Protection Regulation (GDPR), Health Insurance Portability and Accountability Act (HIPAA) and My Health Record (MHR). However, continually evolving state-of-the-art techniques in machine learning, data analytics and hacking are making it even more difficult to completely protect individual's / patient's privacy. In this article, we have systematically examined various secondary uses of EHR with the aim to highlight how these secondary uses effect patients' privacy. Secondly, we have critically analyzed GDPR and highlighted possible areas of improvement, considering escalating use of technology and different secondary uses of EHR.

\end{abstract}


\end{frontmatter}



\section{Introduction} 

Clinical data is generated in the form of ongoing patient diagnostic services. These services usually take place in hospitals, clinics or laboratories through different clinical trials (via medical imaging or doctors prescriptions) or through wireless body area network using wearable sensors \cite{al2017survey}. All of these sources produce huge amount of clinical data world wide and its volume is experiencing an exponential growth. It is estimated that clinical data will swell up to 2314 Exabytes by 2020 from a figure of 153 Exabytes in 2013 with an annual growth rate of 48 percent \cite{pramanik2019healthcare}.

In most of the countries (especially developing countries), data generated during routine clinical practices is stored manually in the form of paper based medical records. This procedure is adopted by the medical practitioners because of ease of handling, lack of understanding or for the purpose of treating more patients in less time. However, this method of storing patient's medical information is not useful for the patients and does not guarantee accurate and timely deliverance of healthcare services. 

Some other problems associated with manual recording of clinical / medical data are:

\begin{enumerate}

\item Paper based medical records can easily be altered or can be lost and may cause sever consequences. 
	
\item Physicians / clinicians can prescribe wrong medications (due to alteration of paper based medical records) or cannot advice right medications during follow up visits without properly knowing past medical records of patients.
	
\item  It is not practical for a person to carry a huge bunch of paper based past medical records during follow up visits or to describe complete medical history to a physician / clinician in case of change of physician or hospital.

\item Reviewing and analyzing paper based records poses cumbersome task for new physicians or medical staff when patients change their physicians or hospital. 

\end{enumerate}

To avoid all of the above described difficulties, automated / electronic online patient information system through which patients' complete medical record is made available to healthcare professionals is required. Such electronic system also serves the purpose of storing patients data for longer time without any alterations and makes it accessible through different locations to support quick decision making processes \cite{maghazil2004comparative}.

 \begin{figure}[!htb]
     \centering
     \includegraphics[scale=0.5]{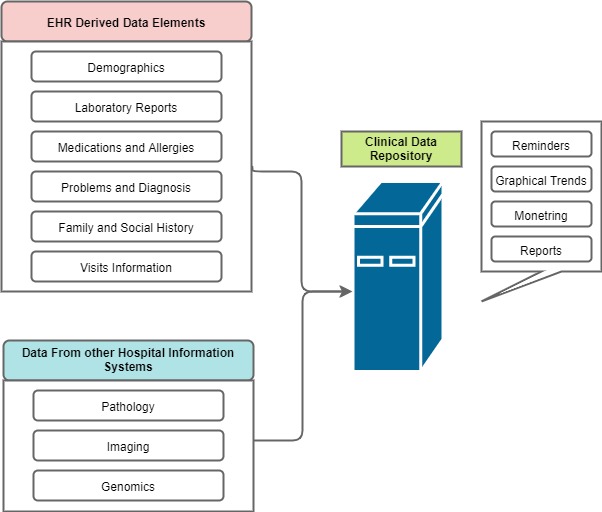}
     \caption{EHR as a clinical data repository}
     \label{Figure_1}
 \end{figure}

Healthcare organizations are now adopting techniques to digitize medical records to overcome challenges (described above) faced by them or by patients while using paper based medical records \cite{ben2015electronic}. With the new technique, patients' clinical data is now stored as Electronic Health Records (EHR). EHR are the patients' computerized health records that contain patients' complete information along with their medical history in a format (refer Figure \ref{Figure_1}) that can be easily shared among different health care providers or can be accessed by them through different linked locations when required \cite{spiranovic2016increasing}.

Adoption of EHR provide range of benefits over the traditional paper based medical record systems. For Example:

\begin{enumerate}

\item EHR are capable of storing structured, coded and electronic patient data all together to form a complete history of patient's health \cite{lobach2007research}. 

\item Electronic data saved as EHR makes a Decision Support System (DSS) for monitoring health outputs to improve health care quality \cite{o2011impact}, where DSS is a tool, usually software based tool, that supports decision making by providing automated analysis of data \cite{dss2015}.

\item EHR system acts as a central database of information for patient documentation and billing, maintaining quality, and supporting patient related sensitive decisions \cite{seymour2012electronic}. 

\item Data saved in EHRs can be accessed through multiple locations simultaneously and also can be shared with different partner organizations conveniently. Thus, making data accessible to the concerned physicians across multiple sites to better provide healthcare services.

\item EHR reduce probability of errors related to medical data analysis as it stores complete medical records and thus lowers overall healthcare cost \cite{menachemi2011benefits}.

\end{enumerate}

With all the above mentioned benefits of using EHR, certain risks factors are also associated with it. The most important issue is the data security and patient's privacy. In case, if  EHR data is leaked or theft from the database, it can be misused (by altering dosage of drugs or treatment procedure etc.) and may cause severe complications or even leads to the patient's death \cite{wang2013research}. It is therefore utmost important to protect patient's information in central database from unauthorized wrong hands. Patient's information may also be theft while it is in transmission to the other linked services over the network.

Information contained in EHR is also used for different secondary purposes (other than patient personal care) such as clinical research, health promotions, clinical audit and clinical governance, national screening and preventive campaigns, audits against national standards, national statistics, planning future services, and resource allocations etc. \cite{teasdale2007secondary} (refer Section \ref{SecUses} for discussion on secondary uses of EHR). For all such uses, patients may not be willing to share their information because often patient share their private health data for their personal care and not for the other / secondary uses. Using patient sensitive information for different secondary purposes without their consent seriously effects their privacy. 

To safe guard patient privacy or personal data, there exist privacy standards in different regions of the world such as General Data Protection Regulation (GDPR) in Europe \cite{regulation2016regulation, albrecht2016gdpr}, Health Insurance Portability and Accountability Act (HIPAA) in the United States (US) \cite{hipaa1, cohen2018hipaa} and My Health Record (MHR) in Australia \cite{hemsley2018legal,MHR1}. These standards provide legislation to protect personal data but with fast paced advancement in data analytics and artificial intelligence \cite{MUNIR2019, KHAN_imavis} poses new challenges for such standards.

Our contributions in this article are following: 

\begin{enumerate}

\item In this study we described various secondary uses of EHR with the aim to highlight how these secondary uses effect patients' privacy, refer Section \ref{SecUses} for discussion on secondary uses of EHR. 

\item In this article we have discussed various issues associated with secondary use of EHR, refer Section \ref{challenge}. Referred section also elaborates on security and privacy issues of using EHR data (Section \ref{secuPrivacy}). 

\item This article systematically analyzed GDPR (recent privacy standard to protect European citizens data) regulation and enlisted its challenges of ensuring that EHR data to be used only for the purpose agreed upon by the patient, refer Section \ref{gdpr}.


\end{enumerate}

Our contributions in this article are oriented toward understanding ethical concerns when dealing with personal data in the era of Artificial Intelligence (AI). Research domain of our contributions (described above) needs more collaborative efforts by research community working in the domain of medicine, computing and law to achieve better insight. Ethical issues arising due to fast proliferation of AI-assisted technologies \cite{Riz2019a} will raise various serious concerns, specially related to privacy of individuals. Due to complex nature of this interdisciplinary research domain it is hard to find literature on the topic and thus, our article is novel as it systematically analyzes uses of sensitive EHR data which, if violated, creates many privacy and ethical concerns.

The rest of the paper is structured as follows: Section \ref{EHR} describes EHR along with their different standards. Section \ref{infoSources} describe information sources of EHR. Section \ref{SecUses} describes use of EHR in various secondary purposes. Section \ref{challenge} presents challenges of using EHR for secondary purposes. Section \ref{gdpr} describes the systematic analysis of GDPR in context with the patients privacy and data security with respect to the secondary uses of EHR. Finally, in section \ref{conclusion} conclusions is presented. 


\section{Electronic Health Records (EHR): Data Sharing} \label{EHR}

EHR is a clinical data repository containing basic patient information such as patient's personal profile, his / her complete family history, laboratory reports, physicians and other medical staff notes etc. Along with this primary information, EHR also contain data form the other hospital information systems such as imaging data from radiology departments, patients genomics data from genetic departments or endoscopic or colonoscopic data  from Gastroenterology departments etc. Figure \ref{Figure_1} illustrates the most important data elements included in the EHR.

\begin{figure}[!htb]
    \centering
    \includegraphics[scale=0.65]{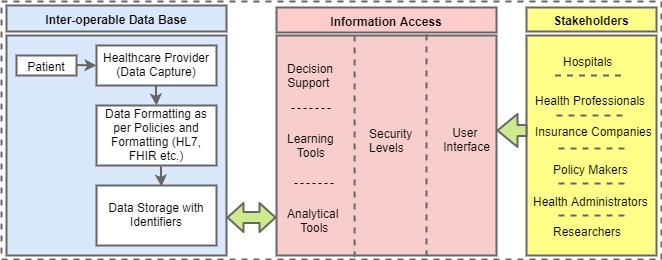}
    \caption{Conceptual overview of EHR system}
    \label{Figure.2}
\end{figure}

EHR also provide functionality of generating reminders for routine screenings and disease reporting, generating graphical trends against various parameters such as blood pressure monitoring, heart beat monitoring, blood glucose level monitoring etc. The same is also shown in Figure \ref{Figure_1}. Such reporting is highly beneficial for patients health and safety especially when patients are in critical condition and their strict monitoring is required.


Conceptually EHR system can be divided into two basic parts \cite{latha2012electronic}. Creation part and the access part (refer Figure \ref{Figure.2}). Creation part is based on the interaction of patient with the healthcare providers. This part explains, how the data from the patient is captured, how it is formatted according to the policies and standard and finally, how the formatted data is stored in an interoperable database. Access part is based on the access of the data stored in EHR by the different authorized users or organization. This part explains how confidential information from EHR can be securely accessed by the authorized users via user friendly interfaces.

 \subsection{EHR standards}

\begin{figure}[!htb]
	\centering
	\includegraphics[scale=0.7]{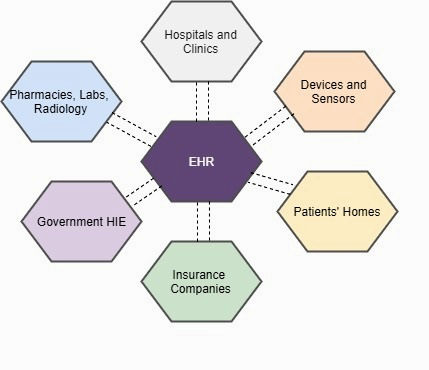}
	\caption{EHR data sharing}
	\label{Figure.3}
\end{figure}

For the effective use of data contained in EHR, it must be shared  through different linked locations such as clinics, hospitals, radiology departments, pharmacies, laboratories and patient homes \cite{hayrinen2008definition} (refer Figure \ref{Figure.3})

Shared data at multiple locations ensures patients solitary care by identifying their basic needs in terms of care, safety, timeliness, and effective monitoring. It also helps medical staff (physician, nurses etc.) to take right actions based on patient conditions. The data usefulness can further be increased if the data contained in EHR is linked with different clinical decision support systems (CDSS). CDSS is automated medical data analysis tool that suggests next steps for treatment and generate alerts by predicting future conditions / trends by analyzing provided data \cite{Kawamoto765}. By this way, the physicians can take sensitive decisions quickly and effectively \cite{castaneda2015clinical}. 

However, without any industry standard for information exchange, it is usually difficult to share and exchange EHR data across multiple sites. The same difficulty was faced by the healthcare organizations to communicate EHR data with each other and with different decisions support systems when there was no industry standard available for health information exchange. It was the main reason behind the slow adoption of EHR system in healthcare organizations even if their adoption was highly beneficial for them \cite{boonstra2010barriers}.

\subsubsection{Health Level Seven (HL7) standard} 

The Health Level Seven (HL7) organization was established in US in March 1987 to develop consistent common standards for Hospital Information System (HIS) \cite{kalra2006electronic}. Afterwords this organization defined HL7-Clinical Document Architecture (HL7 CDA) as EHR messaging standard for easy integration, interchange, sharing and retrieval of information across different clinical information systems. The HL7 standard allows different healthcare organization to share and exchange patient information via encoded data exchange. It provides a common syntax of information for different clinical information systems to share information (contained in EHR) conveniently \cite{seymour2012electronic}. 

The HL7 CDA Framework 1.0 release, became an American National Standards Institute (ANSI) approved HL7 standard in November 2000 \cite{dolin2001hl7}. After release of first version, version 2 and version 3 releases were also made available with some new standards and modifications \cite{dolin2006hl7}. HL7 CDA is a markup for specifying composition and semantics of data ingredients of EHR such as a discharge report, admission summary, progress and procedure reports and to exchange them with various stakeholders. It is an absolute object document that may hold clinical data in various formats such as text, image, sound, or other multimedia content.  Extensible Markup Language (XML) is used to encode the HL7 CDA clinical documents, which then can be exchanged in form of HL7 messages or using other transport solutions.

An HL7 CDA message consists of a header and a body. Header contains information regarding patient, source (provider) and the authentication of the message. On the other hand, the body of the message includes organized clinical reports i.e. lab, radiology, Magnetic Resonance Imaging (MRI), Computed Tomography (CT) scan, ultrasound etc.

\subsubsection{Fast Healthcare Interoperability Resources (FHIR)}

In order to improve inter interoperability and exchange of information, HL7 released different version time to time. In 1988, HL7 version 2 was released to enhance and streamline information exchange mechanisms / procedures, that can be used by different departments across hospitals \cite{ benson2016hl7}. However, different limitations were exposed in this version such as difficult implementation process, having number of optional segments and above all lack of proper representation that is capable enough to identify techniques for exchanging messages and interfaces \cite{beeler1998hl7}. To overcome the shortcomings of version 2, version 3 was developed in the year 1995 . Although, HL7 version 3 resolved much of the problems of previous version, it could not resolve the incompatibility issue raised because of variety of sub versions \cite{al2013hl7}. In order to further improve HL7 standards, another novel interoperability standard i.e. Fast Healthcare Interoperability Resources (FHIR) was initiated in the year 2011 \cite{bender2013hl7} by HL7 organization. FHIR standards are very simple to adapt, possess scalability and are robust in nature. These standards have potential capabilities of supporting work flows in small devices like mobile phones \cite{sharma2019hl} 

\section{Electronic health record information sources} \label{infoSources}

Adoption of EHR is beneficial both for patients, physicians and healthcare providers. It improves overall healthcare quality, omits paperwork, reduce medical errors and increase work efficiency as well as reduce overall healthcare cost \cite{atreja2008using}.  
\par
\begin{figure}[!htb]
    \centering
    \includegraphics[scale=0.6]{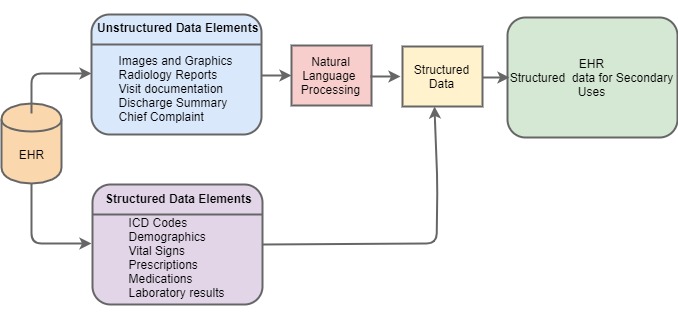}
    \caption{Unstructured and structured data elements of EHR}
    \label{Figure.4}
\end{figure}

Beside patients personal care, EHR data is also used for different secondary purposes(refer section\ref{SecUses} fro secondary uses of EHR). However, functionality of EHR data for secondary uses has been limited because of non uniform data components across EHR. Non uniformity in data elements exists because of the fact that during daily clinical practices, EHR data is often recorded in free text and unstructured format. Therefore, EHR contains structured and unstructured sets of information.  Figure \ref{Figure.4} elaborates more on the structured and unstructured  data components of EHR. As shown in  Figure \ref{Figure.4}, structured data includes laboratory results, vital signs, prescriptions, medications and International Classification of Diseases (ICD) codes whereas, unstructured data includes narrative information (free text) such as images and graphics, radiology reports, visit notes, discharge summary, chief complaint etc.

Figure \ref{Figure.4} depicts that the major portion of EHR data is consisted of unstructured elements. Such data elements are not represented in any standard coding scheme such as ICD codes, therefore, their retrieval, reporting and  aggregation is not easy like structured data using commonly available database tools \cite {atreja2008using}. 
It is therefore required to converted unstructured data elements into structured data in order to make it equally useful for secondary uses. 
\par
One of the method to convert unstructured data into structured data is manually reviewing EHR by the experts using text charts or data abstraction methods \cite{lin2013application}. However, these methods are time consuming and not reliable to capture all the structured information. Furthermore, it is beyond the capacity of human being to review clinical data of EHR in large volume. 
Natural language processing (NLP) has been found a great help for the researchers to extract structured clinical data from unstructured data elements \cite{liao2015development, kreimeyer2017natural}. It is an 
artificial intelligence domain of computer science that uses computers to manipulate unstructured data such as narrative text in form of clinical notes or speech data \cite{wong2018natural}. 
\par
Mostly, NLP uses statistical (probabilistic) Machine Learning (ML) models to derive language data from large volume of free text data. These models use text data to identify common patterns and associations in the data \. NLP based ML models give meanings to words and phrases in text and converts unstructured data elements of EHR into structured codes. In short, NLP captures unstructured data elements of EHR, analyze the data elements with respect to their grammatical structures, obtain meanings from grammatical structures, and finally summerizes information to make it useful.

\section{Secondary uses of EHR} 
\label{SecUses}

One of the contribution of this study, as described above, is systematic analysis of various secondary uses of EHR data with the aim to highlight how these secondary uses effect patient's privacy. This section discusses different popular secondary uses of EHR data. Section \ref{challenge} elaborates on challenges associated with the secondary use of EHR data while, Section \ref{secuPrivacy} focuses on privacy and security challenges of EHR data.

\subsection{Clinical Research}

The basic purpose of clinical research is to use EHR for design and execution of clinical trials for new medicines \cite{coorevits2013electronic}. Health related issues are directly proportional to the population. Currently, healthcare organizations, hospitals, laboratories are facing shortage of trained medical staff due to various reasons. One of the possible solution to tackle different medical ailments is to discover new drugs with better results or new techniques and robust strategies. All such activities require clinical research to be conducted. Thus, clinical research holds pivotal role in tackling some of the hard pressed medical issues.

\begin{table}[!htb]
\caption{Different Domains of Clinical Research}
    \centering
        \begin{tabularx}{\textwidth}{|l|X|}
    \hline
        \textbf{Domains of clinical research} 		& \textbf{Examples}\\
        \hline
        Hypothesis Generation &  To understand the people response on introducing new drugs. \\
        \hline
        Epidemiology & To find causes of disease in community of people.\\
        \hline
       Drug utilization & To figure out the use of medicines and to determine the frequency.\\
       \hline
       Patient recruitment & To raise the awareness of clinical trials.\\ 
        \hline
       Health Technology Assessments & Evaluating large number of research publications on a topic of interest and generating highly consolidated information for policy makers and health care providers \\
          \hline
       Comparative Effectiveness &  Obtaining real world evidences from the analysis of real world data generated through routine clinical practices to help decision makers for making effective decisions.\\
         \hline
       Pragmatic  Trails &  Observing patients' treatment and their outcomes in real world situations to provide on ground real information to the decision makers so that they can make effective decisions for the enhancement pf quality of care.\\
        \hline
    \end{tabularx}
    \label{Table.1}
\end{table}


Some of the other areas where clinical research is required are: 

\begin{enumerate}

\item Prediction of diseases based on patients present data \cite{xiao2018deep}. 
	
\item Study of drug behaviors with different diseases or different patients i.e. study on antibiotics \cite{willyard2017drug}. 

\item Developing vaccines for the prevention of diseases before it attack \cite{spicknall2018review}.

\end{enumerate}

Other than the areas mentioned above, there exists multiple domains (refer Table \ref{Table.1}) where clinical research is essential to overcome the existing problems of the medical world and to ensure high quality of healthcare delivery to the patients.

In the domain of clinical research, EHR is an essential part because it is a basic information source and a possible way of exchanging clinical information with different stakeholders. Based on this exchange of information, various health statistics are developed and decisions are made. For example, based on data collected word wide, World Health Organization (WHO) publishes various reports time to time for public awareness and for the authorities knowledge to understand the current trends and future needs related to particular diseases \cite{world2019global,world2017global} .

Table \ref{Table.2} lists possible information sources available in EHR that can help in successfully carrying out clinical research in different domains. 

\begin{table}[!htb]
\caption{Different information sources available in EHR that can help in carrying out clinical Research}
    \centering
		\begin{tabularx}{\textwidth}{|p{2.5cm}|p{6cm}|X|}
    \hline
       \textbf{Data Sources} &  \textbf{Explanation} & \textbf{Possible areas of Clinical Research} \\
			&   & \\
        \hline
        Demographics &  It includes patient's basic information such as
        name, age, gender, date of birth, address, contact, allergies, past medical history and diagnosis & Data analysis, community based research, age related research,  disease surveillance, and all other epidemiological human population studies.\\
        \hline
        Daily Habits (Risky Behaviors) & Using tobacco, alcohol, and other sedative drugs. & Cancer research, chronic drug usage implications on health, mental illness,  psychological disorders. \\
        \hline
       Facts and Monitoring & Weight, height, blood pressure, blood sugar, heart beat. & Hypertension research, Body Mass Index (BMI) based research, diabetic research, early childhood growth studies and cognitive outcomes.\\
       \hline
       Laboratory Data & Complete Blood Count, Prothrombin Time, Basic Metabolic Panel, Comprehensive Metabolic Panel, Lipid Profile, Liver Functioning Test, Thyroid Stimulating Hormone, Hemoglobin A1C, Urinalysis, Microbiological Culture with antibiotic resistance tests and others & To investigate the origin of disease, study of communicable and non communicable diseases as well as blood disorders.\\ 
        \hline
       Various Encounters Data & Human population or, hyperlipidemia, diabetes, anxiety and obesity, allergies, reflux esophagitis, respiratory problems, depressive disorder, asthma, nail fungus, urinary tract  kidney failure, migraine & Research of all non communicable diseases.\\
        \hline
       Special tests \& Procedures &  Appendectomy, Electrocardiogram, Biopsies, Angiographies, Therapies & Special investigative tests for the advance research on non communicable diseases identification and control.\\  
        \hline
        Imaging  &   Magnetic Resonance Imaging, Ultra Sound, Computerized Axial Tomography,  Positron Emission Tomography etc. & Cancer Research, identification and monitoring of Congenital anomalies  diseases in unborn babies, bone fractures and tumors, can be used to monitor response of tumors to chemotherapy or radiations.\\
          \hline
       
    \end{tabularx}
    
    \label{Table.2}
\end{table}

Table \ref{Table.2} shows that EHR contains enough information to carry out clinical research in different domains. Successful utilization of this information for research purpose requires development of new and emerging research infrastructures capable of exchanging information based on latest published standards. However, when data is shared across different healthcare organizations it raises different security and privacy concerns. These concerns are discussed in Section \ref{secuPrivacy}.


\subsection{Public Health Surveillance}

Another secondary use of EHR is Public Health Surveillance (PHS). PHS is a process of collecting, analyzing and interpreting data related to a specific disease for administrating and assessing public health on the whole \cite{teutsch2000principles}. PHS particularly investigates those diseases, which harm or may tends to harm a large population and grow in community like epidemic diseases. Its main functions include collection of facts about a particular disease, risk factors of its spread  and interpretation and analysis of the collected facts for controlling the disease to prevent public from its severe effects.
One of the example of PHS is the surveillance of Dengue outbreak in Pakistan that has been reported in \cite{ahmad2018surveillance}. Dengue is a viral disease which causes high fever in patients and spreads in people 
because of the bite of a particular Aedes aegypti mosquito. Recently, it has affected around 40\% population of the world. Pakistan is one of the most affected country from it. There are several other examples of PHS world wide like reported in \cite{schwartz2017surveillance, mace2018malaria}.

As discussed above, PHS is a common practice word wide but mostly in third world countries it is performed using manual procedures \cite{khan2007dengue, anvikar2016epidemiology}. In these methods, data about disease for surveillance purpose is collected using traditional methods. For example, physicians prescription records are gathered either from patients or from hospitals and clinics  or through public surveys \cite{shah2010surveillance}, similarly data form other departments of the hospitals (laboratories, radiology departments, emergency departments etc) is also collected manually by visiting the logs of these departments' databases. The collected data is then cross communicated between public health staff and health protecting agencies via telephonic and fax communication networks. Collected data is stored on papers manually and manual procedure is used to analyze the stored data \cite{siswoyo2008ewors}. This method of PHS is time consuming, requires large manpower and needs huge efforts to record, store and analyze the data. It is also not a reliable method as there are chances of errors due to manual handling of data. Inaccurate and uncertain outcomes are possible based on the collection and inspection of manually collected and stored data \cite{sips2017automated}. Such traditional methods are not suitable for the confirmation of certain disease, understanding its severity, its transmission risks, and the spread of other linked diseases.

\begin{figure}[!htb]
    \centering
    \includegraphics[scale=0.6]{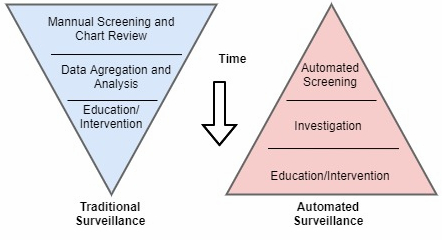}
    \caption{Traditional v/s automated surveillance}
    \label{Figure.5}
\end{figure}

With more effective way, surveillance of diseases can be performed by actively monitoring patients EHR. As EHR are rich in variety of data, the summary generated by analyzed data is provided to public health agencies for prevention and control of diseases. Health surveillance by EHR provides the glance of health status of the community, which promotes the quality of healthcare. It tracks the key diseases, with more effective way than manual procedures. Use of EHR provides the opportunity to automate the PHS. It is an effective way of preventing outbreaks by discovering utmost danger cases irrespective of merely reacting to outbreaks \cite{atreja2008opportunities}. Figure \ref{Figure.5} elaborates the effectiveness of EHR based automated surveillance against the traditional manual surveillance systems. 

During traditional surveillance, the most of the time is utilized on manual screenings and reviewing charts and less time is saved for the actual intervention. On the other hand automated PHS assisted by EHR data is time efficient in analyzing data. The same is shown in Figure \ref{Figure.5}.

Use of EHR for public health surveillance has proved to be effective in developed countries such as United Kingdom (UK), United States (US), France, Norway, Canada and Australia \cite{keck2013influenza}. In these countries, local health departments have diverted their manual surveillance system towards EHR based electronic surveillance system. This practice has advanced the functionality of PHS \cite{birkhead2015uses}. Developing nations have also initiated adoption of EHRs for PHS to robustly analyze data and take actions, if required \cite{odero2007innovative}. Thus, it is an important need of the present day automated surveillance systems to use data from EHR. For example,  Integrated Disease Surveillance and Response System (IDSRS) requires data to be obtained from patient medical records \cite{abdullah2019dengue, onyebujoh2016integrating}. It is therefore necessary for the hospitals and the other health related agencies to adapt EHR so that automated surveillance can be done effectively.


\subsection{Clinical audit and quality assurance}

\begin{figure}[h]
    \centering
    \includegraphics[scale=0.6]{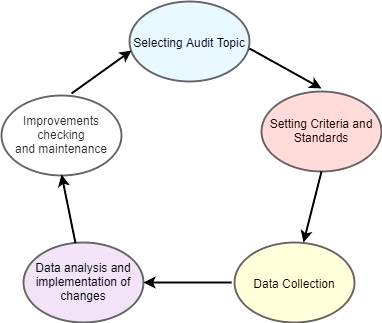}
    \caption{Clinical audit as a cyclic process}
    \label{Figure.6}
\end{figure}

The aim of clinical audit is to enhance patient care via rigorous analysis of care provided against benchmark standards \cite{burgess2011new}. Clinical audit is a systematic way of settling standards, analyzing data based on standards, performing actions to meet settled standards and executing proper monitoring to sustain the standards. Clinical audit is a cyclic process (refer Figure \ref{Figure.6}) that contains different stages to be followed for the achievement of best practices in clinical practices.  

Standards settled for the clinical audits require not only to be obeyed by the medical staff (doctors, nurses, midwives, therapists, etc.) but also by the healthcare organizations like hospitals, clinics, nursing homes, ambulatory surgical centers, autonomous laboratories, radiology units, collection units etc. Clinical audit focuses on broadly accepted methods to improve over all healthcare quality. For example, organizational development, information management and  statistics evaluation are the key functions of clinical audits.

The role of EHR is very important in clinical audits as it provides detailed and accurate information to the auditors. Using EHR for clinical audits gives convenience to the auditors to perform the clinical audits as compared to use of traditional clinical data for audit \cite{esposito2014clinical}. Figure \ref{Figure.6} shows that the data collection and data analysis are the important parts of the clinical audit. In order to perform quality clinical audits, the clinical data must be easily available as well as the available data must be reliable to perform clinical audit. EHR conveniently provide data to the auditors from multiple access points to perform clinical audits to better provide the quality of the care to the patients.

\section{Challenges associated with secondary use of EHRs} \label{challenge}

Primarily, EHR data is collected for patient's individual care and administrative billing purposes. Using this data for different secondary purposes (as elaborated in Section \ref{SecUses}) is always challenging \cite{bayley2013challenges, lobach2007research}. It is because priorities and settings of primary and secondary uses are different. The quality of data collected for the primary purpose cannot be same as the quality of data collected for secondary uses.  For example, data collected for clinical research needs much more care and attention during collection than the data gathered during routine clinical practices in the form of EHR. The quality of collected clinical data is a serious concern of the researchers. Due to this reason, with respect to reuse of clinical data, the authors of \cite{van1991use} suggested that the data must be used for its primary purpose only.

Following are different factors of concern that affect the quality of clinical data collected through EHR.

\subsection{Correctness}

Correctness refers to the accuracy of the collected data that is directly linked with its initial documentation (how the data was collected, recorded and stored). EHR data is collected through routine clinical practices during which the clinicians priority is to collect the patients data according to their own point of interest and according to different administrative needs but not according to their various secondary uses (refer Section \ref{SecUses}). The chances of errors are obvious in this case. According to the study presented in \cite{hogan1997accuracy}, data accuracy collected through EHR ranges between 44\% to 100\%. 

Errors in EHR may lead to different outcomes, if their data is used for various secondary purposes. Errors include:

\begin{enumerate}

\item Inaccurate predictions by clinical researchers
\item Degradation in health standards and statistics as data analyzed was error prone.
\item False health surveillance results that may lead to unforeseen medical emergency. 

\end{enumerate}

Improvement in accuracy of EHR is essential to make it equally beneficial for primary as well as secondary uses.

\subsection{ Incompleteness}

Another factor that effects quality of clinical data is related to the completeness of the EHR. usually, EHR do not contain complete patient history. It is because patients do not always trust single healthcare organization and may visit several such organization to get sense of satisfaction.  Study conducted in \cite{bourgeois2010patients} showed that out of 1.1 million adult patients, 31\%  visited two or more hospitals, whereas, one percent patients visited five or more hospitals for acute care during five year period of their study.
  
Patients also miss follow up visits suggested by the physicians or sometime due to the perfunctory of the concerned medical staff (who records patient related data), incomplete records are stored in EHR. A Study was conducted at Columbia University on 3068 pancreatic cancer patients out of which only 48\% patients had complete pathology records, while, the rest had incomplete records about the disease \cite{botsis2010secondary}.

EHR data is also considered incomplete for secondary uses because of the data ``locked-up'' condition.  Locked-up condition means records have details regarding patient but it is not present in the coded portion of the record or in other words data present in EHR is structured and unstructured (already described in Section \ref{infoSources}). Structured data is in the format that can be easily processed by the computers. On the other hand unstructured data mostly requires Natural Language Processing (for hand written prescriptions) technique to be applied to make it structured (detail is provided in Section \ref{infoSources}) and processable by computers \cite{kho2013practical, ramakrishnan2010mining}.


\subsection{Inconsistency}

EHR data is handled by various individuals and at different locations. Multiple persons are involved in entering, storing and processing the data, therefore, data contains several definitions. Most of the data is present without mentioning proper units as units are often remembered by the medical staff and they can understand language written by each other. On the other hand for the non concerned person (who want to use the data for secondary purpose), it may be highly difficult to interpret the data without specified units. Involvement of different individuals in preparation and processing of EHR leads to an inconsistent form of data. It means, data present in EHR is not uniform. In such non uniform data, it is often difficult to relate assessments of different practitioners (because the assessment of different clinicians is often different). Secondly, data inconsistency also arises due to the fact that the data is collected with different tools at different locations, which may be time varying (data coding regulations and system abilities may change with time) \cite{bayley2013challenges}. Inconsistent data may lead to erroneous data analysis and wrong results. Therefore, such inconsistent data is not useful for secondary use.

\subsection{Security and privacy challenges} \label {secuPrivacy}

EHR based clinical data provides many advantages over manual paper based medical records. It is cost effective, improves overall healthcare quality and above all can be easily accessed through different linked locations. All such advantages motivate health providing agencies and medical practitioners to adopt EHR based system. However, adoption of EHR and its data processing introduces several privacy and security issues. Especially, when this data is used for secondary purposes (refer Section \ref{SecUses} for discussion on secondary uses of EHR). In the next subsections, security and privacy challenges related to secondary uses of EHR have been separately discussed.

\subsubsection{Security challenges} \label{secHa}

EHR contains patients personal and highly confidential data in the form of physicians' personal notes, neuroimaging data \cite{arxiv2020, 211a}, X-rays, ultrasounds as well as lab reports. This data may include lab results of HIV and other sexually transmitted diseases \cite{JULIEN20153}, mental disorders \cite{129ab}, personality disorders \cite{BATEMAN2015735}, contagious diseases as well as doctors sensitive comments about patient mental illness or personality disorders etc. All such data is stored in hospital's local database (each hospital may have its own local electronic database), which is connected across other hospitals or health providers databases via internet or wireless connections for sharing purposes. Transfer of such confidential data over the internet creates several security risks. It provides a chance to hackers and other harmful attackers to access the data and use it for their own purpose and to effect large number of patients \cite{jamia019}. In case of patients monitored at home, the data from patients is collected through a distributed network of sensors. Securing such data is another big challenge because there are greater chances of spying and skimming \cite{meingast2006security}.

With the passage of time healthcare technologies are extending and new technologies are being introduced to provide instant help to the patients and to enhance healthcare quality. For example, different smart devices monitor health (with the general purpose devices or wearable sensors) and prescribe medications as well as provide telemedicine technology for delivering remote care \cite{dimitrov2016medical}. Patients now can easily access healthcare facilities by integrating their mobile phones with telemedicine and telehealth services with the help of simple mobile applications \cite{WEINSTEIN2014183}. As the technology in the healthcare is continually evolving, its inter connectivity is also evolving. With the help of interconnected network, patients information is made broadly available to the relevant organizations and staff to provide quality healthcare. Exchange of patient information over the large inter connected network is beneficial in many ways but has increased existing security risks.

Cyber security is a technology that  safeguards computer networks and information contained in them from different cyber attacks \cite{arxiv2019ab}. In case  of healthcare data, cyber security technology needs to be robust and strong as the healthcare sector presents lucrative avenue to cyber criminals to attack and get hold of very sensitive data to gain large financial benefits. Unfortunately, the results from study conducted by Kruse et al. \cite{kruse2017cybersecurity} concluded after analyzing data from thirty one (31) articles that the healthcare industry lags behinds in security as compared to the organizations working in other domains i.e. education, business sector, entertainment etc.


The prime reason for criminals to target healthcare data is to get financial gain. Criminals sell valuable data taken from EHR to the ``darkweb'' \cite{dWeb} (darkweb refer to the content on the web that is not indexed by search engine and thus remains hidden from the general public) and achieve high financial gain. For the criminals, EHR data is more informative than credit cards because it contains various fixed identifiers and important financial information that is extremely worthy in black markets. Fixed identifiers of of EHR data can not be reset like the once in credit cards. Such identifiers in EHR are best information sources for the criminals to get easy access to the patients bank accounts for getting  loans or to capture their passports and other important documents (property, insurance etc.) \cite{kruse2017cybersecurity}. For example, recent new article published story about theft of EHR data (20,000 record) from North Carolina-based Catawba Valley Medical Center. Stolen data contained patient names, dates of birth, medical data, health insurance information and social security numbers \cite{newsEHR} .

It is worth mentioning here to explain that the security of the healthcare data is not only today's concern rather it was the concern before the emergence of the EHR \cite{IThealth}. Data security was well studied before the EHR came into existence (paper based patients records were needed to be safeguarded within the premises of a hospital and not on large scales) but with the adoption of EHR multiple gateways opened for accessing patients' information remotely. Furthermore, The patients EHR contains more detailed information all together in a single source as compared to the previous paper based medical records, which were distributed among different departments of the hospitals. With the adoption of EHR it is now easy for the criminals to attack millions of people at a time and to stole their valuable information (because EHR are interconnected with numerous networks. In case of paper based records it was not possible to stole millions of patients records at a time).

In short, adoption of EHR not only provided the range of benefits but also introduced potential risks of cyber attacks. Healthcare organizations spend more on increasing their integration but do not spend much on their system protection. In order to gain patients trust and to give them satisfaction regarding their data safety, the healthcare providers have to think about developing  robust practical standards and solutions with particular healthcare / EHR needs.


\subsubsection{Privacy challenges} \label{PriChall}

Privacy is defined as ``right to be left alone'' or to keep away from public domain \cite{bookPrivacy}. United nations general assembly (UNGA) declared privacy as a fundamental human right in its universal declaration of human rights. However, in this digital era the term privacy has become subjective and is interpreted and implemented differently by each state or country \cite{Kayaalp2018}. Such ambiguities are sometimes exploited for different reasons, for example EHR data is used to gain financial benefits \cite{bookFinan} or for different secondary purposes, refer Section \ref{SecUses} for discussion on secondary uses of EHR.

As mentioned above, EHR data contains several security risks especially when the information contained in them is shared with different stakeholders over the interconnected networks. Other than security issues, there are certain privacy concerns linked with exchange and sharing of EHR data. These privacy concerns are usually raised due to the fact that when the patients data (which was recorded for the purpose of patient individual care) is being shared or linked without consent or knowledge of particular individual. Usually consent of an individual is necessary for sharing of data but ambiguity arises when different healthcare organizations have different perspective on question of ``who owns the data?''. Is data belongs to patient, his / her physician, health insurance organization, healthcare organization, social security agency or is it jointly owned by all \cite{IThealth, NAP9750}?

Breach of data can happen due to various reasons, refer Section \ref{secHa}, which has many ethical repercussions. For example, disclosing patient's sensitive private information such as sexually transmitted diseases or mental illness in the public domain can negatively impacts individual's reputation. In extreme cases such individuals can face social boycott as people start avoiding an individual if they knew that he / she has sexual transmitted disease like HIV, chlamydia etc \cite{boyc}. Secondly, person's status in the society is seriously affected if his / her mental illness is disclosed to the public \cite{ment2005}.  Another dimension to this issue is financial impact on individual's life as medical insurance companies usually calculates premium / cost of insurance based on medical history and life events. In such cases insurance companies can increase their premium \cite{bookAmer, ABBAS2015}.

The privacy of clinical data has been subject to a lot of research and it has been difficult to determine how much of the data belongs to the patient and how much of it may belong to healthcare organizations and whether the consent of the owner of data is needed, in case the data is to be used for the research purpose \cite{Richter2019, NAP9750, jama.294}. Privacy of patients can be affected when his / her data is used for clinical research or secondary use, refer Section \ref{SecUses} for discussion on secondary uses of EHR. For example, blood sample given by the patient is stored in a laboratory and after carrying out requested analysis the same sample is analyzed again for the purpose of clinical research. Even though the sample is returned back to the laboratory without any damage, still it violated data privacy because by this way the patient control over his / her data was lost \cite{NAP9750, Richter2019}.

In research conducted by Bovenberg and Almeida \cite{Bovenberg2019} referred to a case of patients versus Myriad Genetics, a molecular diagnostic company. The case was about four US cancer patients who wanted to have full access to their genomic data. Myriad claimed that patients were provided with all the information that was necessary to be included in their reports and additional data was not part of the medical record set. Patients, however claimed that the additional data was acquired from their lab sample, hence they have the right over data and only they should decide what happens to their data.

In order to protect sensitive data many patients try to conceal their sensitive information. It is because of the lack of confidence on the system's security retaining their data. It also shows mistrust of patients' on medical staff (doctors, nurses and the others) because patients think that they might disclose their confidential information to public that may create embarrassment for them in society \cite{SADAN200141}. Some events have happened in the past because of which patients have become more sensitive in disclosing their private information. For example, in 2013, one of the medical technician of a US hospital was found guilty in selling patients medical information \cite{fbiReport}. Similarly, a hospital in the US informed his 34000 patients that their medical information has been lost from their agent \cite{ozair2015ethical}. Due to all such incidents, patient don't feel confident in disclosing their information even to the physicians. Hiding facts and information from the physicians and the medical staff can lead to treatment failure. Thus, such challenges may have severe consequences for patients, healthcare providers and even for the governments. 

It is highly recommended from policy makers, leaders and related authorities to discuss privacy and security concerns of EHR data (database storage policies or its sharing policies and paradigms) and formulate policies to address these concerns. There are some existing policies, which need to be revised or reformulated according to the present day era, an era of data analytics, big data and artificial intelligence.

\section{General Data Protection Regulation (GDPR) and its challenges} \label{gdpr}

In order to protect patients personal sensitive data from different security threats and privacy violations, in some regions of the world, data protection regulation have been enforced by the authorities. The most popular data protection regulations are General Data Protection Regulation (GDPR) \cite{regulation2016regulation}, Health  Insurance  Portability  and  Accountability Act (HIPAA) \cite{hipaa1} and My Health Record (MHR) \cite{MHR1}. In this study we have focused only on GDPR and have critically analyzed it in terms of how it protects patient privacy and enforces data security. 



After years of discussions, drafting, negotiations and efforts, in April 2016 GDPR was passed by European Union.  On 25 May 2018, the European Parliament and Council of the European Union both with their combined efforts enforced the GDPR 2016/679 \cite{politou2018backups}. Since then, professionals, citizens and authorities across the Europe and beyond are strictly bound to the legal regimes imposed by GDPR. It is an exhaustive document of legislation that addresses challenges of data protection of  personal data. The aim of GDPR is to control and improve handling and processing of personal data particularly of European citizens. It oversee every aspect of citizens personal data handling and has recommended to impose heavy penalties for non compliance that may include prosecution of any organization in the world that is found guilty of privacy breach or misusing European citizens data \cite{sirur2018we}.

GDPR is not only beneficial for the citizens but also for the organizations as it gives citizens confidence to share their data with the organizations when required. It also boosts organizations business and help them in their smooth running without any hurdle of acquiring citizens data (without trust citizens usually do not share their data when required by the organizations, refer Section \ref{PriChall} for discussion on mistrust between data provider and data handler). Even with all these obvious advantages, organizations in the past were rigid to adapt (at present they are forced to adapt)  privacy regulations imposed by GDPR \cite{gruschka2018privacy}. This is due to the fact that enterprises and organizations were facing challenges in implementing these regulations \cite{tankard2016gdpr}. The organizations were already complying with the regulations imposed by the European Data Protection Directive (EDPD) of 1995 \cite{Directive} and were not prepared for the new changes or possibly there was a lack of awareness for the new requirements raised by the GDPR. Another issue with the implementation of GDPR was financial needs, human resource requirements as well as proper training of the employees to understand the GDPR regulations \cite{tikkinen2018eu}.

GDPR defines six main data protection principles (other data protection principles further clarify them or further enhance them) that organizations (healthcare organizations) have to comply with when processing European citizens personal data \cite{goddard2017eu}. 


Each of these principle is briefly explained below with implications on EHR data.

\begin{enumerate}

\item \textbf{Lawfulness, fairness and transparency} (Article 5(1)(a)): This article states that citizens personal data must be processed lawfully, fairly and transparently. Lawful processing of data is further defined in Article 6, which states that in order to process personal data lawfully, it is necessary for the data controllers to set out / obey one of the following conditions. In this section the term ``data controller'' is used multiple times and in the context of this study this term refers to healthcare organizations which records and stores/hold personal data.

\begin{itemize}

	\item  ``The data subject must be given consent (Article 6(1)(a))''. 
	\item  ``Processing is necessary for the performance of a contract to which the data subject is party (Article 6(1)(b))''.
	\item  ``Processing is necessary for compliance with the law (Article 6(1)(c))''.
	\item  ``Processing is necessary to protect vital interest of the data subject (Article 6(1)(d))''.
	\item  ``Processing is necessary for the performance of a task carried out in the public interest (Article 6(1)(e)''.
	\item  ``Processing is necessary for a legitimate interest of the controller or third party (Article 6(1)(f)''.
	
\end{itemize}

In order to process personal data lawfully, all the clauses of the Article 6 (mentioned above) are important to be followed by the data controllers but the most pertinent clause of the article 6 in the context of EHR data is  6(1)(a) that relates to the processing of personal data with the consent of the person whose data is being used. However, based on the employer-employee or physician-patient relationships, where one party (physician in our case) is in power and processes other party’s personal data, consent is not a proper legal basis to be relayed upon \cite{taylor2019insight}. This is due to the fact that data protection regulation requires consent should be genuinely free without any pressure / intimidation. It can only be possible if the patients have freedom in giving their consent or not and have a choice to withdraw their consent at any point of time without any detriment as easy as they gave it.


\item \textbf{Purpose limitations (Article 5(1)(b))}: Purpose limitations bounds organizations (healthcare organizations) and individuals to collect personal data only for a specific, explicit and legitimate purpose and the data must be used for achieving that purpose only. Data purpose must be clearly defined before its collection and it should not be further 
processed in a way that is incompatible with the original defined purpose(s).

\item \textbf{Data minimization (Article 5(1)(c))}: In order to use personal data, it must be limited to its primary purpose only. It must not be collected more than its need.

\item \textbf{Accuracy(Article 5(1)(d))}: In dealing with the citizens personal data it must be responsibly dealt for example, if the data needs updation and inaccurate or incomplete data elements need to be removed, all must be done with high accuracy.

\item\textbf{Storage limitations (Article 5(1)(e))}: Storage limitations refer to the fact that personal data must be deleted after it has been used and no longer further needed. It means data should be collected with a proper predefined time-line and it must be removed after the the time-line is reached.

\item \textbf{Integrity and Confidentiality (Article 5(1)(f))}: It is the entire responsibility of the individuals or organizations (who want to process citizens personal data) to ensure the safe processing of data and to protect it from unauthorized use. During processing, data must be safe from any accidental loss, damage or demolition and it must be protected against any unlawful use.  

\end{enumerate}

\subsection{Critical analysis of GDPR with reference to EHR data}

If analyzed critically, clauses (b-f) of Article 5 have contradictory nature in the context of EHR data concepts. The regulations mentioned in these clauses (such as data minimization, purpose limitation) limits the quantity of data collection and enforce its deletion soon after the purpose has been achieved. On the other hand,  healthcare organizations encourages collecting more and more amount of data and to save it for longer period of time for the purpose of detailed analysis, mining and predictions \cite{tene2012big}, as discussed in Section \ref{SecUses}.

Article 25 further enhances the ideas presented in Article 5 by defining privacy by design i.e. ``The controller must implement appropriate technical and organizational measures for ensuring that, by default, only personal data which are necessary for each specific purpose of the processing are processed''. Although, this Article enhances protection of personal data by demanding privacy by design form the controllers but it is difficult to implement because of its broader definition and due to the requirement of additional implementation cost and resources. Furthermore, privacy by design can show rigid behavior with time (like the other embedded technical solutions)  because of not updating its measures frequently \cite{bincoletto2019data}.

It has already been described in this study (refer Section \ref{challenge}) that the healthcare data is one of the most vulnerable data in terms of security threats, therefore needs special attention for protection during processing. Article 9 of GDPR defines the processing of such especial categories of data, which required additional protections in processing such as genetic data, biometric data, healthcare data etc. Article 9 imposes additional obligations and provides more restrictive legal basis for processing health related sensitive data. The recommendation of this article is to obtain explicit consent of collecting and processing sensitive personal data. Although, explicit consent of data processing is required in processing any type of personal data (Refer article 6(1)(a) mentioned above) but in case of processing healthcare data, obtaining consent is usually difficult, specially for secondary purpose, refer Section \ref{SecUses} for secondary uses of EHR data. Obtaining explicit consent for every secondary use is a time consuming, costly as well as an exhausting process \cite{article9}. There has been a great debate on obtaining specific consent in literature. The conclusive outcome of all such debates is to shift specific consent into a broader consent of data processing that covers range of its future uses (such as secondary uses of EHR) \cite{harle2019does}.   


At present, most of the patients are not aware (or do not want to be aware) about what happens to their data once it has been taken from them and also they do not know about the data processing procedures undertaken by the healthcare providers. According to Spiekermann et al.  \cite{spiekermann2015challenges}, if the individuals knew about today's  healthcare business model and how third parties use personal private data, they would be surprised and feel betrayed. Obviously, under such circumstances, obtaining broad consent is not logical.

Article 32 of GDPR defines security of processing of personal data. According to it, to process and maintain security of personal data pseudonymisation should be performed \cite{regulation2016regulation}. Pseudonymisation is a technique to ensure that individual won't be identified through personal data (personal data includes direct and indirect identifiers that can identify a person for example, name , ID number, location, contact information (Article 4)) \cite{regulation2016regulation}. The process is to replace main characteristic of an individual with randomly generated indicators. The information regarding identification must be stored separately \cite{voigt2017scope}. Even if  pseudonymisation technique is applied, it is possible to re-identify individuals by combining different data sets \cite{zarsky2016incompatible}. Re-identification pull downs the illusion of privacy policies, which are promised by technologists.  Lawmakers should re-evaluate law and consider weakness of pseudonymisation \cite{ohm2009broken}.

Other than the regulations described above, one of the most controversial regulation is the ``Right to be Forgotten'' (Article 17). This article imposes obligation of erasure of one's personal data on the controllers. It gives right to the users to erase their data any time from all the available places from where they want as per their request. According to concept of healthcare data where decision support and predictive systems are being made by archiving the patients' personal data (consider case of public health surveillance or clinical research, refer Section \ref{SecUses}), this article create huge controversy because logically no more backups or archives of data would be applicable by the organizations.

\section{Conclusion}  \label{conclusion}

The objective of this research article is to provide overview of EHR and its various secondary uses, how such uses effect individuals privacy and whether existing privacy regulation i.e. GDPR overcome these privacy challenges. Article began with overview of EHR, its data sources that contribute in making EHR and advantages of using it. Then, different standards for sharing EHR data i.e. HL7 and FHIR are discussed. Secondly, thorough analysis of various secondary uses of EHR with the aim to highlight how these secondary uses effect patients' privacy is presented. In the last article critically examined GDPR and highlighted possible areas of improvement, considering escalating use of technology and different secondary uses of EHR. 

Presented article outlined various secondary uses of EHR to give readers an idea that how effectively EHR data can be used in different  domains such as clinical research, public health surveillance and clinical audits to provide effective, timely and quality healthcare facilities to the patients, refer Section \ref{SecUses}. In order to use EHR data for secondary purposes more effectively,  challenges associated with the secondary uses of EHR have also been described to make readers well aware of the EHR data challenges when using it for secondary purposes.

In the present technological era, adoption of EHR has positively impacted healthcare services. With the help of seamless data sharing an individual can avail instant healthcare services at his / her location of preference. However, with evolving technology, risks of data security and compromise of privacy have also been significantly increased. EHR data contains highly personal and sensitive information i.e. ID / social security number, bank details, family information and medical history. Unauthorized access to EHR information can have devastating financial and social impact on individual if such sensitive information is leaked in public sphere. In this article different ethical and privacy issues arising from EHR data leak are discussed with detail in  Section \ref{secuPrivacy}. In the referred section, data security and patients privacy risks related to the secondary uses of EHR especially when EHR data is transmitted through network and shared \& exchanged with multiple stake holders are critically studied.

There exists privacy regulations such as GDPR, HIPPA and MHR to protect patients privacy and data security when EHR data is used for secondary purposes and transferred \& exchanged with multiple concerned through different linked locations. However, there is a need to critically examine such regulations to analyze them for calculating their effectiveness in terms of safeguarding personal data as per present era needs. There is also a need to highlight the challenges of such regulations to further improve their effectiveness in safeguarding personal data from the potential cyber attacks and to cope with the technological advancements of cyber attacks. Study presented in this article focused only on GDPR. GDPR's most relevant clauses (related to privacy and clinical data security) are studied in perspective of secondary use of EHR, refer Section \ref{gdpr}. Our purpose is to highlight possible improvements areas in GDPR regulations to make it more effective in protecting privacy and data security and to make it robust against escalating AI-assisted techniques in data analytics and cyber attacks.





\end{document}